\documentclass{article}

\PassOptionsToPackage{numbers,compress}{natbib}
\usepackage[preprint]{neurips_2025}

\usepackage[utf8]{inputenc} 
\usepackage[T1]{fontenc}    

\DeclareUnicodeCharacter{03BC}{$\mu$}

\usepackage{hyperref}       
\usepackage{url}            
\usepackage{booktabs}       
\usepackage{amsfonts}       
\usepackage{amsmath}        
\usepackage{nicefrac}       
\usepackage{microtype}      
\usepackage{xcolor}         
\usepackage{graphicx}       
\usepackage{float}          
\graphicspath{{figures/}}   

\title{Beyond Black Boxes: Enhancing Interpretability of Transformers Trained on Neural Data}

\usepackage{authblk}

\author[1]{Laurence Freeman}
\author[1]{Philip Shamash}
\author[2]{Vinam Arora}
\author[1]{Caswell Barry}
\author[1]{Tiago Branco}
\author[2]{Eva Dyer}

\affil[1]{University College London}
\affil[2]{Georgia Tech}

\begin{document}

\maketitle

\begin{abstract}
    Transformer models have become state-of-the-art in decoding stimuli and behavior from neural activity, significantly advancing neuroscience research. Yet greater transparency in their decision-making processes would substantially enhance their utility in scientific and clinical contexts. Sparse autoencoders (SAEs) offer a promising solution by producing hidden units that respond selectively to specific variables, enhancing interpretability. Here, we introduce SAEs into a neural decoding framework by augmenting a transformer trained to predict visual stimuli from calcium imaging in the mouse visual cortex. The enhancement of the transformer model with an SAE preserved its original performance while yielding hidden units that selectively responded to interpretable features, such as stimulus orientation and genetic background. Furthermore, ablating units associated with a given variable impaired the model’s ability to process that variable, revealing how specific internal representations support downstream computations. Together, these results demonstrate that integrating SAEs with transformers combines the power of modern deep learning with the interpretability essential for scientific understanding and clinical translation.
\end{abstract}

\section{Introduction}

Transformer architectures have recently achieved state-of-the-art performance in modeling neural population activity, demonstrating their ability to capture long-range temporal dependencies and integrate diverse cell-type information \citep{vaswani2017attention, AzabouEtAl2023b}. These models hold promise for building general-purpose representations of brain function, but they also inherit a central challenge of modern deep learning: their internal mechanisms are often opaque, with little visibility into what has been learned or how decisions are made. This “black box” nature has raised concerns in scientific domains where interpretability is essential for hypothesis generation and causal understanding \citep{lipton2018mythos, rudin2019stop}. In neuroscience and clinical applications—where model insights must often be mapped back onto biological mechanisms or used to guide decisions—the lack of interpretability can be a critical barrier to trust and adoption \citep{samek2017explainable, caruana2015intelligible}. Neural recordings further complicate this issue, as they lack a predefined vocabulary or token structure, unlike language, making it difficult to identify consistent, interpretable patterns across sessions, subjects, or brain regions in an unsupervised manner \citep{khaligh2014deep}. Mechanistic interpretability aims to close this gap by linking internal model activations to human-understandable features.

A common approach uses \emph{linear classifier probes} to assess whether hidden representations linearly encode target features \citep{AlainBengio2017}. While this reveals the presence of information, linear-only methods do not attempt to decompose the underlying features into separable interpretable features. Sparse autoencoders (SAEs) offer an alternative: by learning an overcomplete dictionary, they extract \emph{monosemantic} features, each aligning with a coherent pattern in the data \citep{cunningham2023sparse}. In natural language, SAEs have decomposed transformer activations into thousands of interpretable single concept features \citep{BrickenEtAl2023}, and related methods have enabled concept unlearning in diffusion models \citep{CywinskiDeja2025}. In molecular biology, training SAEs on protein language model embeddings has uncovered sequence motifs and functional domains, demonstrating the generality of the method across modalities \citep{AdamsEtAl2025}. Although first applied to neuroscience over two decades ago, sparse autoencoders have recently gained prominence through their successful use in language model interpretability, but have yet to be applied to transformers trained on neural data \cite{BrickenEtAl2023, olshausen1997sparse}.

Despite these advances, standard SAE formulations require careful tuning of sparsity penalties and can mix unrelated features. \emph{TopK}‐SAEs address this by explicitly retaining only the $k$ largest activations per sample \citep{MakhzaniFrey2013}, yielding highly selective, sparse latents that facilitate downstream interpretation. In this work, we integrate a TopK‐SAE bottleneck into the POYO+ transformer \citep{AzabouEtAl2025}, which was trained on one of the largest-scale neural datasets currently available, covering hundreds of recording sessions across 256 mice, from the Allen Brain Observatory \citep{deVriesEtAl2020}. POYO+ was selected not only for its state-of-the-art performance but also because the dataset on which it was trained targets the mouse visual cortex, a system with a widely studied and relatively well-understood functional architecture. In particular, early visual areas such as the primary visual cortex (V1) are known to exhibit simple and structured receptive fields, including orientation- and spatial frequency-selective responses that can be linearly approximated \citep{hubel1962receptive,ringach2002spatial,carandini2005early}. These characteristics make the visual cortex an ideal domain for interpreting learned representations in neural networks.

We show that (1) the SAE preserves decoding performance on downstream performance tasks, (2) individual SAE latents exhibit sharply tuned “receptive fields” linking units to specific stimuli and cell types, and (3) targeted ablations of SAE units yield causal, feature‐specific effects on prediction accuracy. Our results pave the way for transparent, mechanistic models in neuroscience.  

\begin{figure}[H]                
  \centering
  \includegraphics[width=1\textwidth]{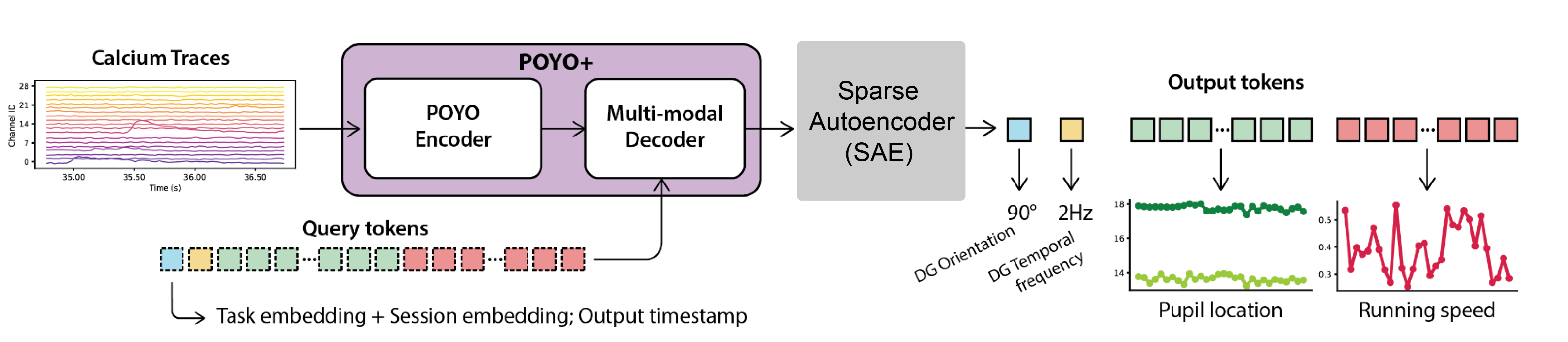}
  \caption{%
    \textbf{Integrating a Sparse Autoencoder into the POYO+ architecture.}  
    Raw calcium fluorescence traces recorded across multiple channels are encoded by POYO+, passed through a Sparse Autoencoder (SAE) in order to constrain the latent residual stream to a sparse representation, reconstructed to the original latent dimensionality, and then linearly projected to predict downstream variables such as pupil location and running speed. This figure was adapted from the POYO+ paper. \cite{AzabouEtAl2025}%
  }
  \label{fig:architecture}
\end{figure}

\section{Methods}

We outline the methodologies employed to analyze and interpret the latent representations derived from the POYO+ transformer decoder. Initially, we detail the procedure for extracting and reshaping POYO+ latents into suitable inputs for a sparse autoencoder (SAE). Subsequently, we describe the introduction of sparsity into the latent space using a TopK-SAE. Finally, we explain our approach for associating SAE latent activations with behavioral conditions, decoding anatomical and genetic labels from latent representations, and performing targeted ablation experiments to assess the causal contributions of specific latent dimensions.

\subsection{Extracting Latents from POYO+}

POYO+ builds on the Perceiver IO framework \cite{jaegle2021perceiver}, by first compressing the full neural time‐series into latent tokens via an encoder cross‐attention “bottleneck,” and then integrating these latents with task query tokens through a final \emph{cross‐attention layer} that produces the task‐conditioned representations used for prediction \cite{AzabouEtAl2025}. We focus our mechanistic analysis on the outputs of this final cross‐attention layer—i.e.\ the moment when distilled neural information and task context have been fused into a compact latent code. At this stage the model has already performed a form of in‐network dimensionality reduction, making these latents a natural target for mechanistic analysis. Concretely, we denote the extracted tokens as $\mathbf{y}_{\ell,k,i}\in\mathbb{R}^{d}$, for recording session $\ell$, task $k$, and time-step $i$, with dimension $d=64$. Aggregating these outputs across $S$ samples and a context window $T$, we construct a tensor
$\mathbf{Y}\in\mathbb{R}^{S \times T \times d}$ this tensor retains temporal and feature dimensions. To facilitate subsequent SAE training, we reshape $\mathbf{Y}$ by collapsing the temporal dimension, obtaining:
\[
  \widetilde{\mathbf{Y}} = \mathrm{reshape}(\mathbf{Y}, (S T) \times d) \in \mathbb{R}^{(S T) \times d}
\]
where each row of $\widetilde{\mathbf{Y}}$ serves as an independent input vector $\mathbf{x}\in\mathbb{R}^{d}$ to the SAE, effectively ignoring auto-regressive properties within a context window. 

\subsection{Introducing sparsity into the transformer latents}

To impose interpretability on the latent representations, we employed a TopK sparse autoencoder (TopK-SAE) with latent dimensionality $n = \gamma d$, where $\gamma > 1$ (typically called an expansion factor). A standard SAE encoder-decoder is:

\begin{equation}
    \mathbf{z} = \mathrm{ReLU}(W_{\mathrm{enc}}(\mathbf{x} - \mathbf{b}_{\mathrm{pre}}) 
    \mathbf{b}_{\mathrm{enc}}),\quad
    \hat{\mathbf{x}} = W_{\mathrm{dec}}\mathbf{z} + \mathbf{b}_{\mathrm{pre}}
\end{equation}

where $W_{\mathrm{enc}}\in\mathbb{R}^{n\times d}$, $W_{\mathrm{dec}}\in\mathbb{R}^{d\times n}$, and bias vectors $\mathbf{b}_{\mathrm{pre}}\in\mathbb{R}^{d}$, $\mathbf{b}_{\mathrm{enc}}\in\mathbb{R}^{n}$. We optimize the reconstruction objective without additional sparsity penalties:

\begin{equation}
\mathcal{L}(\mathbf{x}) = \|\mathbf{x}-\hat{\mathbf{x}}\|^2
\end{equation}

by using the TopK variant. In which $x$ represents the baseline POYO+ latents and $\hat{x}$ represents the sparsely encoded SAE latents and $\mathcal{L}(X)$ forms the loss function. Sparsity is enforced explicitly by retaining only the $k$ largest activations:

\begin{equation}
  \mathbf{z} = \mathrm{TopK}(W_{\mathrm{enc}}(\mathbf{x} - \mathbf{b}_{\mathrm{pre}}))
\end{equation}

This method ensures highly selective embeddings of the POYO+ latents. After the SAE has been trained, we inject the SAE as an additional module in the architecture between the decoder and the linear projection layer that maps the decoder outputs to the behavioral features (Figure 1). Thus we are left with two sets of latents after inference: (1) baseline latents which are not passed through the SAE and are used as a comparison to the (2) SAE latents which form a sparsely encoded set of POYO+ latents.

\subsection{Allen Brain Observatory Visual Coding Data}

We conducted our interpretability analyses on the POYO+ model \cite{AzabouEtAl2025}, which was previously trained on the Allen Brain Observatory Visual Coding dataset \cite{deVriesEtAl2020}—a large-scale collection of in-vivo calcium imaging data recorded from awake mice. This dataset includes GCaMP6f-expressing neurons sampled across six visual cortical areas, multiple cortical layers, and 13 genetically defined Cre-lines covering both excitatory and inhibitory neuronal subtypes. For our analyses, we filtered the dataset to include only recordings from the \textbf{primary visual cortex (VISp)}, a well-characterized region where neurons exhibit interpretable receptive field properties such as orientation and temporal frequency tuning. This anatomical focus enabled clearer evaluation of latent feature selectivity and model representations. Among the 13 Cre-lines, we focussed on two—\texttt{Vip} and \texttt{Sst},—which label distinct inhibitory neuron subtypes. These lines exhibited the highest decoding performance in our preliminary evaluations and reflect biologically distinct interneuron classes with well-known roles in cortical computation. 

Although POYO+ was trained on five stimulus paradigms (drifting gratings, static gratings, natural scenes, natural movies, and sparse noise), we focused exclusively on the \textbf{drifting gratings} task for all interpretability experiments. This stimulus presents full-field sinusoidal gratings drifting in one of eight directions (0°, 45°, …, 315°) at one of five temporal frequencies (1, 2, 4, 8, 15 Hz), yielding a structured two-dimensional stimulus space. Its low-dimensional, interpretable design makes it ideal for probing model representations and performing targeted ablations. Full dataset structure and Cre-line details are provided in Appendix~\ref{appendix:data}.

\subsection{Linking Latent Activations to Downstream Tasks}
\label{sec:linking-latents}

To interpret and quantify how each learned latent dimension relates to behaviorally relevant features, we mapped the unlabeled SAE activations onto the two decodable properties of the drifting‐gratings task: orientation (8 classes) and temporal frequency (5 classes).  This mapping is essential because, although the SAE discovers a compact representation of the neural data, by itself it does not indicate which latent dimensions encode orientation tuning versus temporal‐frequency tuning.  By linking each latent to stimulus labels, we can visualize its “receptive field” in stimulus space and compare how well different architectures (baseline transformer vs.\ SAE‐augmented) capture task‐relevant structure. 

\medskip

\noindent\textbf{Procedure.} Let $Z_{i}(n)$ be the activation of latent unit $i \in \{1,\dots,D\}$ on trial $n$, and let $y^{\mathrm{ori}}(n)$ and $y^{\mathrm{temp}}(n)$ denote the orientation and temporal frequency logits (of dimensions 8 and 5) from the output of the model. We compute class probabilities as $P^{\mathrm{ori}}_{n,j} = \sigma(y^{\mathrm{ori}}(n))_j$, where $\sigma$ denotes the softmax function. We then form a joint stimulus–activation distribution by weighting each trial’s class probabilities by the magnitude of the corresponding latent activation:
\begin{equation}
    P^{\mathrm{joint}}_{i,j} 
    \;=\;
    \sum_{n=1}^{N}
      P^{\mathrm{temp}}_{n,i}\,
      P^{\mathrm{ori}}_{n,j}\,
      \frac{\exp\!\bigl(\lvert Z_{i}(n)\rvert\bigr)}
           {\sum_{n'=1}^{N}\!\exp\!\bigl(\lvert Z_{i}(n')\rvert\bigr)}.
\end{equation}

For trials with high confidence (i.e.\ $\max_{j}P^{\mathrm{ori}}_{n,j}>0.5$ or $\max_{i}P^{\mathrm{temp}}_{n,i}>0.5$), we average the absolute activations and renormalize across $(i,j)$ to produce an \emph{activation map}
\[
A_{i}\;\in\;\mathbb{R}^{5\times 8},
\]
where $A_{i}(i,j)$ captures how strongly latent $i$ responds to the temporal‐frequency bin $i$ and orientation bin $j$.  We interpret each $A_{i}$ as the “receptive field” of that latent, allowing direct comparison across all $D=192$ SAE dimensions (d=64 * expansion = 3).

\subsection{Decoding Cre‐Line and Brain‐Region Labels}

To assess whether model representations capture unsupervised biological features, we decoded genetic Cre-line identity and anatomical brain region from both the baseline transformer and sparse autoencoder (SAE) latent spaces. Specifically, we extracted the latent vector corresponding to each neural sample during the drifting-gratings task and trained a multinomial logistic regression model to classify either Cre-line or brain region labels, using session-wise cross-validation to prevent data leakage. Each sample was filtered to the primary visual cortex, and all classes were downsampled to equal size to ensure balanced evaluation. We evaluated performance using overall accuracy, per-class accuracy, and confusion matrices. Full preprocessing and decoding details, including class balancing, cross-validation, and visualization procedures, are provided in Appendix~\ref{appendix:cre_decoding}.

\subsection{Ablation Experiments on Latent Directions}

To probe the causal role of individual latent dimensions—and to enable precise feature steering for both trained and emergent properties—we performed targeted ablation experiments by selectively zeroing out subsets of the SAE latents.  Such controlled interventions are crucial for aligning model behavior with desired criteria (e.g.\ human values or clinical safety) since they let us remove information about known task features (orientation, temporal frequency) as well as unexpected “semantic” concepts (e.g.\ Cre‐line identity or brain region) that the model learned without supervision. \noindent\textbf{Selecting ablation targets:} For trained features (orientation, temporal frequency), we selected the class combination with the highest activation in their respective bins from the activation maps $A_i$ (see Section~\ref{sec:linking-latents}). For emergent features (Cre-line, region), we fit multinomial logistic regressions on each latent $Z_i$ and selected those with statistically significant positive coefficients ($p<0.01$) for each class (See Appendix \ref{appendix:decoding_stats}).

\noindent\textbf{Ablation procedure.}  For a chosen set of indices $\mathcal{I}\subset\{1,\dots,D\}$, we construct a binary mask vector $\mathbf{m}\in\{0,1\}^D$ with
\[
m_i = 
\begin{cases}
0, & i\in\mathcal{I},\\
1, & i\notin\mathcal{I},
\end{cases}
\]
and apply it element‐wise to the SAE latent activation vector $\mathbf{z}\in\mathbb{R}^D$:
\begin{equation}
  \tilde{\mathbf{z}} = \mathbf{m}\odot \mathbf{z}\,.
\end{equation}
This ablated code $\tilde{\mathbf{z}}$ is then passed through the remaining POYO+ prediction pipeline to measure its impact on downstream decoding performance. By comparing performance before and after ablation on both trained and emergent tasks, we (1) verify that the targeted latents are indeed causally necessary for encoding the feature of interest, and (2) demonstrate our ability to surgically steer model representations—an essential capability for safely deploying large models in clinical neural‐interface settings or other applications requiring fine‐grained control over what information informs predictions.  

\section{Results}

\subsection{Incorporating a Sparse Autoencoder Preserves Decoding Performance}

\begin{figure}[!htbp]  
  \centering
  \includegraphics[width=1\textwidth]{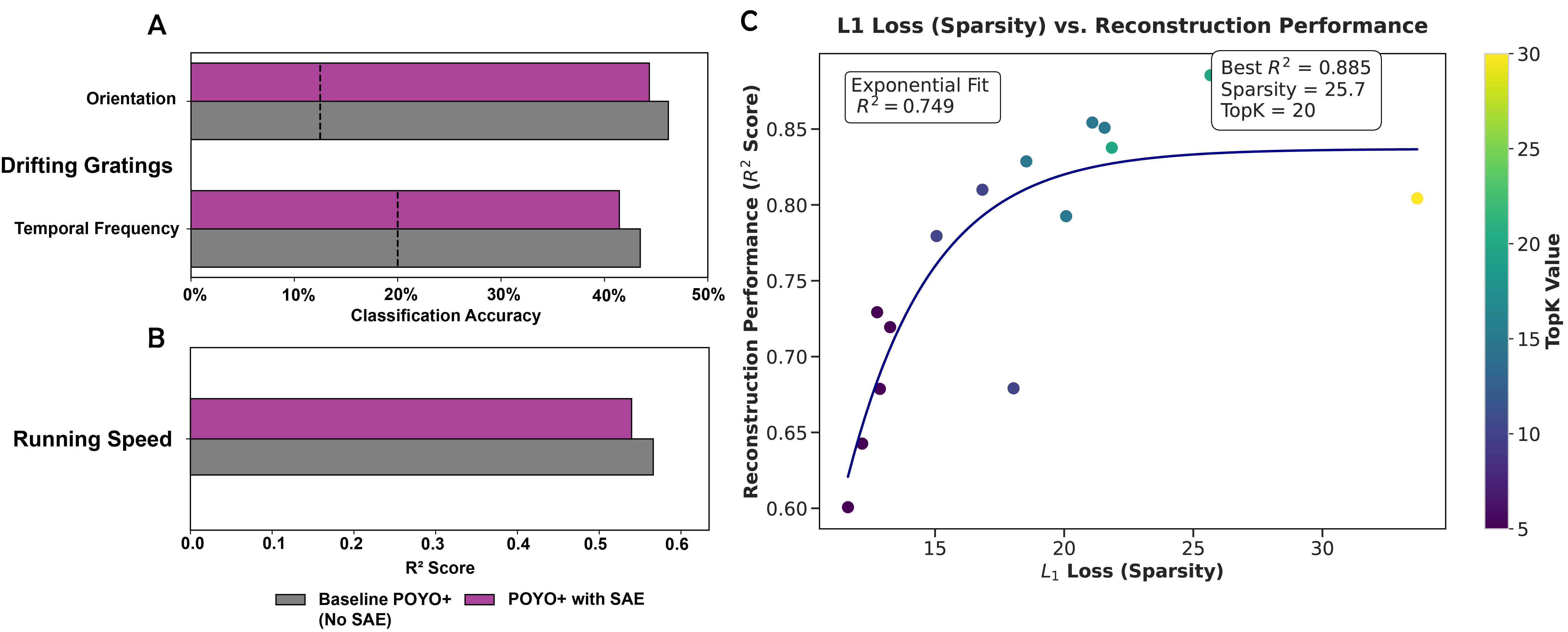}
  \caption{%
    \textbf{Decoding and Reconstruction Performance Comparison between Baseline and Sparse Autoencoder Models.} 
    (\textbf{A}) Classification accuracy for orientation and temporal frequency decoding tasks.
    (\textbf{B}) Prediction of running speed (regression task) also demonstrates slightly suppressed performance when incorporating the SAE model compared to baseline latents. Dotted lines represent chance levels.
    (\textbf{C}) Sparsity analysis of SAE latent dimensions showing reconstruction performance ($R^2$) against latent sparsity ($L_1$ norm). Points represent different TopK configurations and the resulting effect on sparsity levels and reconstruction performance.
  }
  \label{fig:performance}
\end{figure}

To determine whether introducing a TopK‐SAE bottleneck impairs behavioral decoding, we compared the original POYO+ model to a variant in which latents are passed through a trained SAE before the output layer (Fig.~\ref{fig:architecture}). We first assessed classification performance on the drifting‐gratings task: baseline POYO+ achieved $\approx45\%$ accuracy on both orientation and temporal‐frequency decoding, and POYO+ with an integrated SAE maintained accuracies above $40\%$, a modest $\approx5\%$ drop that remains well above chance (12.5\% and 20\%, respectively; Fig.~\ref{fig:performance}A). Next, in predicting continuous running speed, the baseline $R^2$ of $\approx0.57$ fell by only $\approx0.05$ after SAE insertion (Fig.~\ref{fig:performance}B). Finally, by sweeping the TopK sparsity parameter from 5 to 30 retained activations, we charted reconstruction performance, quantified as the coefficient of determination $R^2$ between baseline latents and their SAE reconstructions, against SAE sparsity, defined as the $\ell_1$ norm (the sum of absolute values) of all SAE activations. The best variant (TopK = 20) achieved $R^2 = 0.885$ while retaining a sum of $25.7$ nonzero activations (Fig.~\ref{fig:performance}C). Together, these results demonstrate that a properly tuned SAE yields highly sparse, interpretable embeddings with minimal impact on both classification and regression performance, laying the foundation for targeted feature‐steering and mechanistic analyses. Figure~\ref{fig:performance}C further illustrates the exponential trade‐off between sparsity and reconstruction fidelity: lower sparsity favors interpretability by reducing the number of active SAE units per sample, but comes at the cost of reduced $R^2$.  Thus, selecting SAE checkpoints at the “elbow” of this curve provides an optimal balance between interpretability and reconstruction performance.

\subsection{Mapping Latent Representations to Drifting‐Gratings Features}

\begin{figure}[!htbp]
  \centering
  \includegraphics[width=1\textwidth]{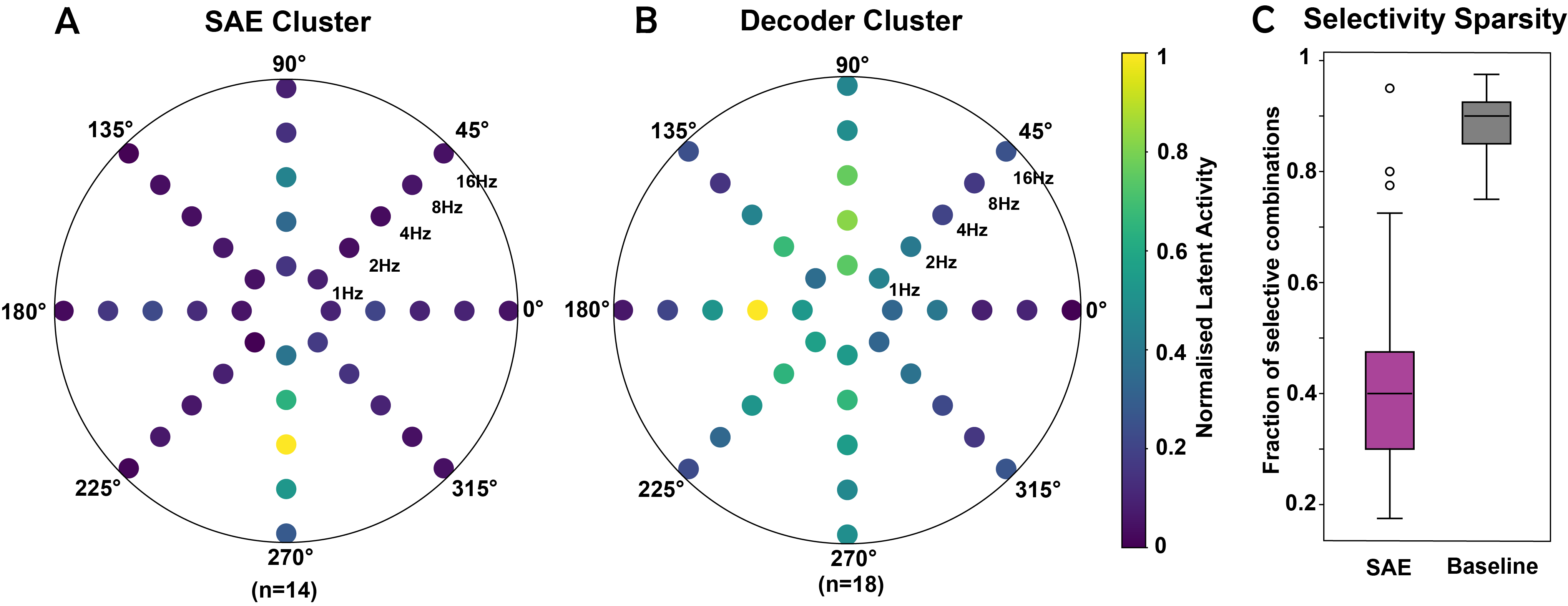}
\caption{\textbf{Sparse Autoencoder Latents Exhibit Increased Task Selectivity Compared to Baseline Latents.} 
(\textbf{A}) Average activation map of 14 SAE latents clustered via UMAP and HDBSCAN, showing selective activation across orientation (angular coordinates) and temporal frequency (radial coordinates). 
(\textbf{B}) Average activation map of 18 baseline latents using the same clustering method, exhibiting broader, less selective activation. 
(\textbf{C}) Quantitative comparison of selectivity sparsity, defined as the fraction of stimulus conditions eliciting an activation. SAE latents display significantly sparser selectivity patterns, enhancing interpretability of downstream representations.
}
  \label{fig:neurips-sparsity}
\end{figure}

Having confirmed that the SAE preserves decoding performance on downstream tasks, we next visualized how its enforced sparsity shapes mappings to the drifting‐gratings task.  We began by clustering latent units according to their trial‐wise activation maps (Section \ref{sec:linking-latents} ): latent activations were first embedded via UMAP and then grouped with HDBSCAN (see Appendix Section \ref{sec:unsupervised_clustering}). For each cluster, we computed the average activation map of the latents over orientation (0°, 45°, …, 315°) and temporal frequency (1–8Hz) and plotted these in polar coordinates (Figure \ref{fig:neurips-sparsity}A–B).  In this representation, angle indicates orientation, radial position indicates frequency, and color reflects normalized response strength. Thus, these plots visualize linking the latent unit activations directly to the behavioral measurements.

An example SAE‐derived cluster of latents (n=14) exhibit highly focused selectivity responding robustly to only orientation–frequency pairs on the 270 degree arm (Fig.~\ref{fig:neurips-sparsity}A). By contrast, a baseline latent cluster (n=18) show broader, more diffuse activation across stimulus space (Fig.~\ref{fig:neurips-sparsity}B). To quantify this difference across every latent, we measured the fraction of orientation–frequency bins exceeding a high‐confidence threshold for each latent, and compared these fractions between SAE and baseline (Fig.~\ref{fig:neurips-sparsity}C). SAE latents occupy significantly fewer bins (median 0.4) than baseline latents (median 0.9), demonstrating that the SAE yields markedly sparser, more selective tuning to drifting‐gratings features. These results confirm that (1) the SAE’s sparsity constraint produces interpretable, behaviorally meaningful latents rather than mere reconstruction artefacts, and (2) these sparse codes form a cleaner foundation for subsequent mechanistic and ablation analyses. 

\subsection{SAE Preserves Cre‐Line Decoding While Enhancing Interpretability}

\begin{figure}[htbp]
  \centering
  \includegraphics[width=\textwidth]{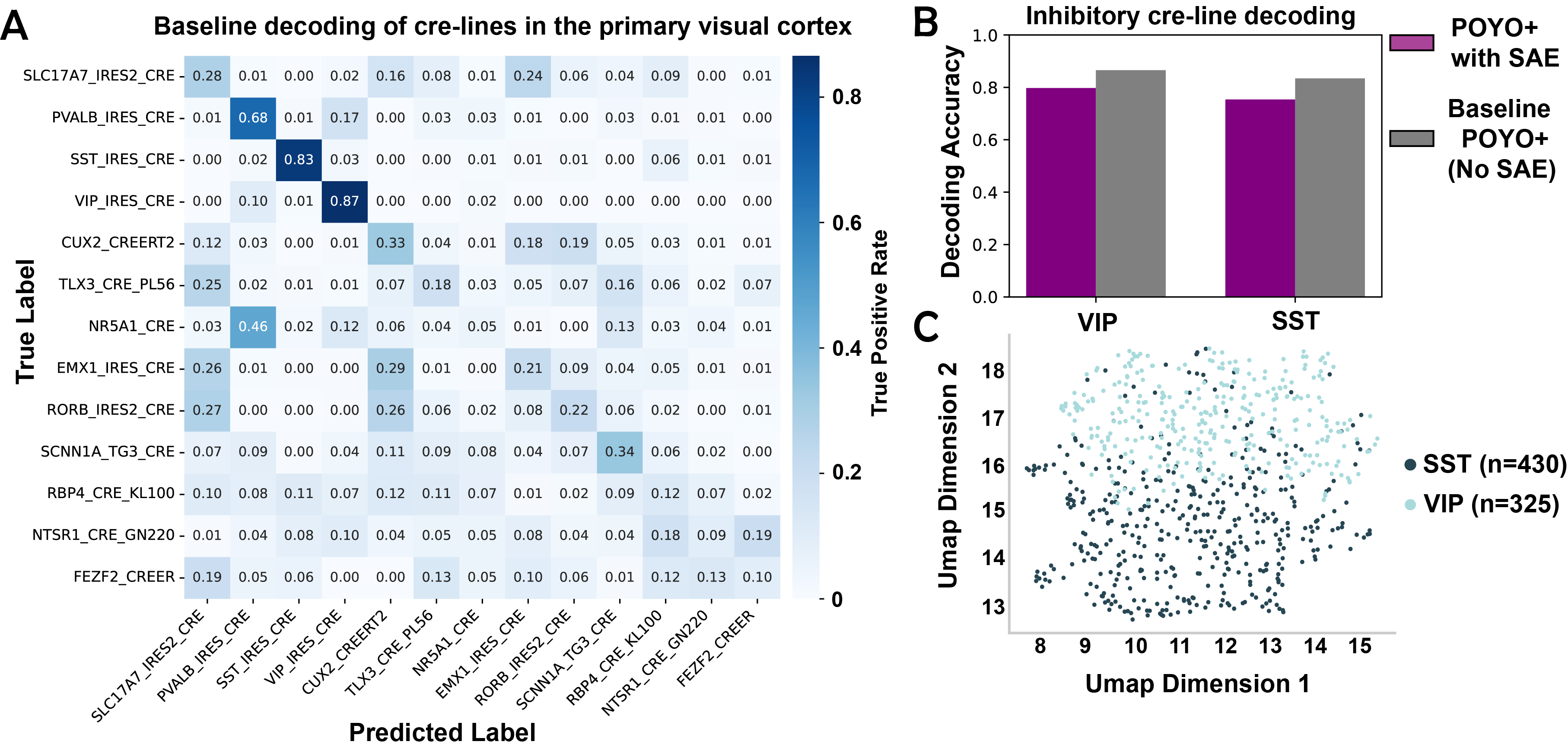}
  \caption{%
    \textbf{SAE Latents Maintain Genetic Cre-line Decoding with Enhanced Interpretability.}
    (\textbf{A}) Confusion matrix showing Cre-line classification accuracy from baseline latents, with strong performance for inhibitory neuron lines SST and VIP using logistic regression.
    (\textbf{B}) Decoding accuracy comparison for SST and VIP Cre-lines between baseline and SAE-transformed latents, showing only a slight performance reduction with SAE insertion.
    (\textbf{C}) UMAP visualization of SAE latents, illustrating clear separation between SST (green) and VIP (orange) Cre-lines, indicating latent disentanglement.
  }
  \label{fig:cre-line-decoding}
\end{figure}

We next asked whether the SAE bottleneck compromises the model’s ability to decode genetic Cre‐line identity—an emergent feature for which the model was never explicitly trained.  Figure~\ref{fig:cre-line-decoding}A shows that logistic regression on baseline latents achieves high Cre‐line classification accuracy.  After inserting the TopK‐SAE (Fig.~\ref{fig:cre-line-decoding}B), decoding performance remains virtually unchanged, with only a modest drop in accuracy across all Cre‐lines.  

Crucially, the SAE also sharpens the alignment between latent units and genetic labels.  UMAP projections of the SAE latents (Fig.~\ref{fig:cre-line-decoding}C) reveal that individual dimensions respond selectively to single Cre‐lines, examples include distinct clusters for SST‐ and VIP‐expressing neurons, whereas baseline latents typically mix multiple Cre‐lines within the same dimension.  This one‐to‐one tuning significantly improves interpretability, making it straightforward to target and ablate Cre‐line–specific representations for causal analyses when attempting to understand what information is encoded within an individual activation function in the network.

\subsection{VIP-Associated SAE Latents Capture Enhanced Directional Selectivity Reflecting Biological Properties}
\begin{figure}[htbp]
  \centering
  \includegraphics[width=\textwidth]{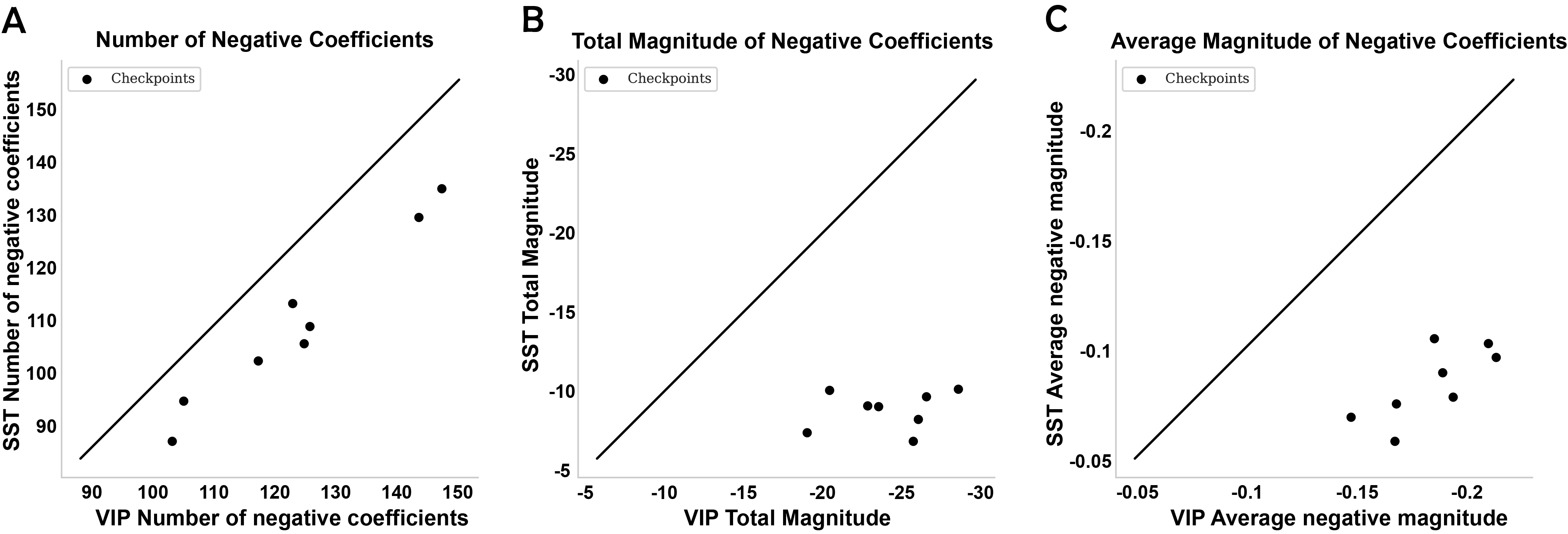}
  \caption{%
    \textbf{VIP-Associated SAE Latents Exhibit Stronger Directional Selectivity Than SST Latents.}
    Logistic regression coefficients linking SAE latents to orientation decoding were analyzed across eight independently trained SAE checkpoints. 
    We compared VIP- and SST-associated latents, focusing on negative coefficients as indicators of directional tuning.
    (A) Number of negative coefficients, with VIP latents consistently showing higher counts.
    (B) Total magnitude of negative coefficients, indicating stronger inhibitory responses for VIP latents.
    (C) Average magnitude per negative coefficient, confirming greater and more consistent directional inhibition in VIP-associated latents. 
    Diagonal lines denote equal values between VIP and SST groups.
  }
  \label{fig:negative_VIP}
\end{figure}

Building on the logistic regression mappings from SAE latents to genetic Cre-lines, we next evaluated whether latents associated with VIP or SST neurons reflect known biological differences in directional selectivity. 
VIP interneurons are known to exhibit strong directional tuning, with inhibitory responses across a broad range of stimulus orientations \cite{MillmanEtAl2020}, whereas SST interneurons display comparatively weaker directional selectivity.
We therefore hypothesized that SAE latents predictive of VIP cells would exhibit a greater number and magnitude of negative coefficients when decoding orientation stimuli compared to SST-associated latents. Figure~\ref{fig:negative_VIP} confirms this prediction.
VIP-associated latents consistently show a higher number of negative orientation coefficients than SST-associated latents across eight independently trained SAEs (Figure~\ref{fig:negative_VIP}A).
Moreover, both the total magnitude (Figure~\ref{fig:negative_VIP}B) and the average magnitude (Figure~\ref{fig:negative_VIP}C) of negative coefficients are elevated for VIP-associated latents, consistent with robust directional inhibition.
The consistency of these findings across checkpoints highlights that SAE latents not only enhance interpretability but also may reliably capture biologically meaningful features of inhibitory neuron subtypes. See Appendix~\ref{appendix:decoding_stats} for details on how significant latents corresponding to orientation and cell types were determined.

\subsection{Ablation Analysis Reveals Causal, Sparse, and Selective Encoding by SAE Latents}

\begin{figure}[htbp]
  \centering
  \includegraphics[width=0.95\textwidth]{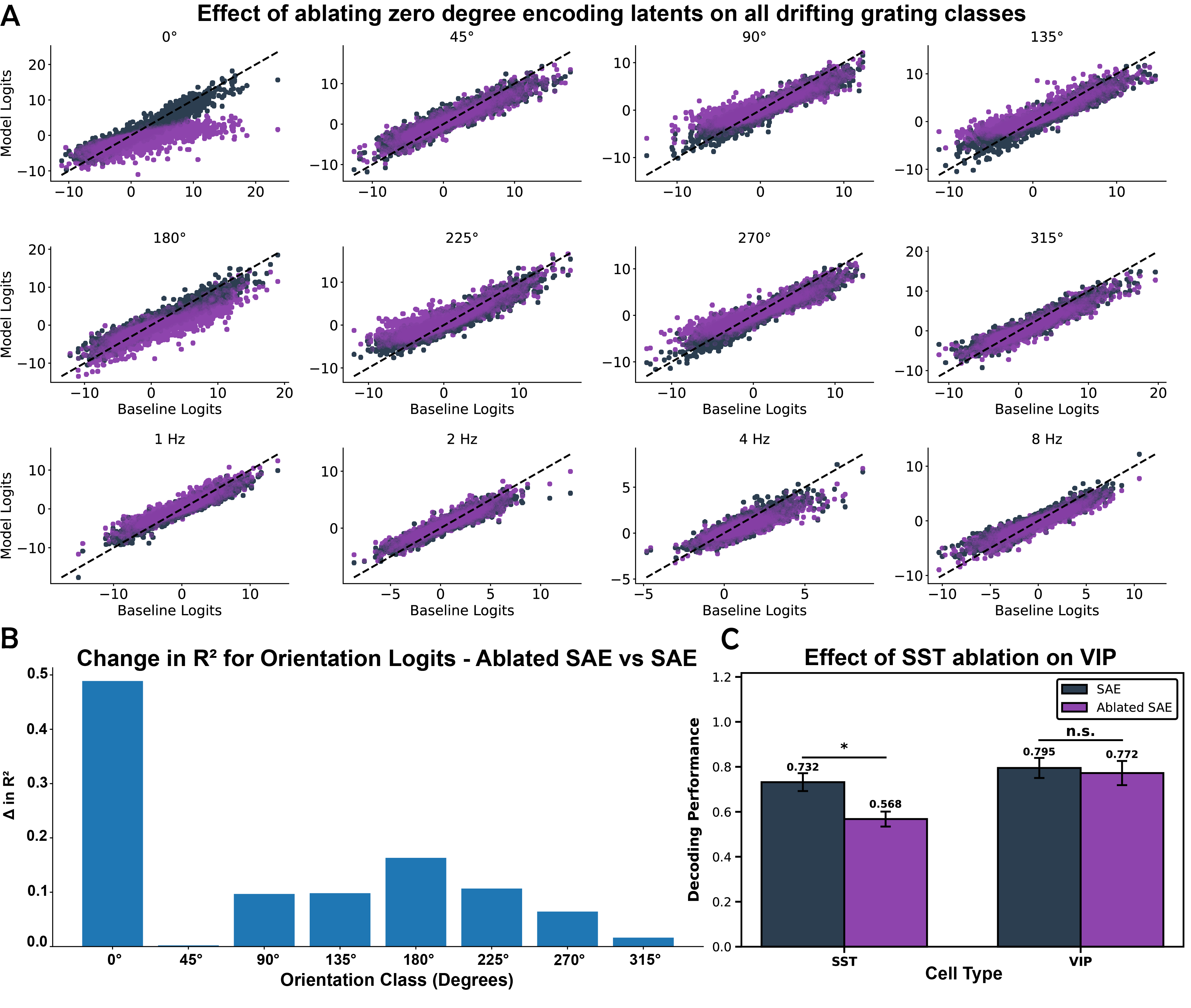}
  \caption{%
    \textbf{Ablation of Specific SAE Latents Demonstrates Causal and Selective Encoding of Task and Emergent Properties.}
    (\textbf{A}) Ablation of SAE latents selectively encoding the $0^\circ$ orientation stimulus across twelve independently trained SAEs. Ablation predominantly impairs decoding accuracy specifically at the $0^\circ$ orientation, while minimally affecting other orientations, confirming highly selective latent representation for task features. (\textbf{B}) A quantification of the change in R2 from the baseline logits to the SAE and ablated SAE models.
    (\textbf{C}) Decoding performance impact from ablating SAE latents strongly associated with SST interneurons. SST decoding performance significantly decreases (* $p<0.05$), while VIP decoding remains unaffected (n.s., not significant). This selective effect is possible due to the sparse encoding by individual SAE latents, a capability unavailable in baseline latents. Error bars represent standard error across cross-validation folds.%
  }
  \label{fig:latent-ablation}
\end{figure}

To validate the causal role of individual SAE latents in encoding task features, we performed targeted ablation analyses. 
In Figure~\ref{fig:latent-ablation}A, ablating clusters of SAE latents selectively tuned to the $0^\circ$ orientation from the drifting grating task resulted in a substantial, specific drop in decoding accuracy for the $0^\circ$ stimulus, with minimal impact on other orientations. 
This demonstrates the precise and sparse feature encoding achieved by individual SAE latents across multiple checkpoints (Figure~\ref{fig:latent-ablation}B).

To further assess the encoding of biological properties, we ablated latents selectively associated with SST interneurons (Figure~\ref{fig:latent-ablation}C). 
This intervention significantly reduced SST Cre-line decoding accuracy (from $0.73$ to $0.56$, $p<0.05$) without affecting VIP decoding, confirming that SAE latents capture SST-specific information selectively and independently. Such targeted, feature-specific ablations are made possible by the sparse disentanglement introduced by the SAE; in contrast, baseline latents, which entangle multiple features, do not support similarly precise interventions.
These results highlight the mechanistic interpretability enabled by sparse autoencoders within transformer architectures trained on neural data.

\section{Discussion}

This work demonstrates that SAEs offer a principled method for mechanistically interpreting transformer models trained on large-scale neural data. By enforcing sparsity in the latent space, we revealed a set of monosemantic latent units, each selectively responsive to biologically interpretable features such as orientation, temporal frequency, or genetic Cre-line identity, without substantially compromising decoding performance. These units not only mirrored known neurobiological tuning properties (e.g., the heightened directional selectivity of VIP interneurons), but also enabled precise targeted ablations to probe causal structure within the model.

The ability to perform such targeted interventions is particularly impactful for two reasons. First, from a neuroscience perspective, it enables conditional independence testing within a high-dimensional learned model: we can ask how downstream representations (e.g., predictions of behavior or genetic identity) change when information about a specific factor is selectively removed from the latent code. This is analogous to classical lesion experiments in neuroscience, but applied in silico to internal representations of deep networks trained on biological data. Such mechanistic dissection of network behavior is essential for validating whether model inferences align with known causal structure in the brain—or for discovering novel structure outright. Second, from an engineering standpoint, this feature-level controllability has immediate implications for safety-critical applications such as brain–computer interfaces (BCIs). Our results show that it is possible to surgically remove latent units encoding undesired or sensitive features—such as genetic identity which may extend to ablating clinical co-variates, while preserving task-relevant performance. This introduces a new pathway for representation-level alignment in neural decoding systems, allowing developers to enforce constraints or safety desiderata on what concepts a model is allowed to use, even if those concepts are not directly trained on. These capabilities arise precisely because the sparse autoencoder enforces a bottleneck that disentangles task-relevant and emergent information into interpretable units. In contrast, baseline transformer latents entangle multiple factors within single dimensions, obscuring interpretability and precluding precise interventions. Thus, by adapting language-model interpretability tools to neuroscience, we demonstrate not only cross-domain generalization of sparse decompositions, but also the beginnings of a systematic toolbox for interrogating and steering the internal computations of black-box models trained on brain data.

Although SAEs enable interpretable latent disentanglement, some limitations remain. First, targeted ablation experiments revealed that silencing SAE latents strongly associated with SST interneurons did not fully abolish SST decoding performance. This partial effect arises because some SAE latents encode mixed information about both SST and VIP cell identities, indicating that intergrating an SAE does not guarantee complete feature orthogonality. Thus, although our approach improves selectivity compared to baseline latents, residual entanglement remains, and achieving perfect disentanglement may require additional architectural or training modifications. Future work could explore sparsity-regularized objectives explicitly penalizing feature overlap or multi-stage approaches to iteratively refine latent specialization.

\section{Appendix}

\subsection{Allen Brain Observatory Visual Coding Dataset}
\label{appendix:data}

The Allen Brain Observatory Visual Coding dataset \cite{deVriesEtAl2020} is the largest publicly available resource of in-vivo two-photon calcium imaging recordings from awake, head-fixed mice. It provides over 500 hours of neural activity data from genetically defined subpopulations of excitatory and inhibitory neurons, offering a highly structured and diverse platform for studying population-level neural coding. Neuronal activity was measured using the genetically encoded calcium indicator GCaMP6f, expressed under the control of 13 Cre-driver lines. These Cre lines target distinct neuronal subtypes:
\begin{itemize}
  \item \textbf{Pan-excitatory lines:} \texttt{Emx1}, \texttt{Slc17a7}
  \item \textbf{Excitatory sub-type lines:} \texttt{Cux2}, \texttt{Rorb}, \texttt{Scnn1a}, \texttt{Nr5a1}, \texttt{Rbp4}, \texttt{Fezf2}, \texttt{Tlx3}, \texttt{Ntsr1}
  \item \textbf{Inhibitory sub-type lines:} \texttt{Vip}, \texttt{Sst}, \texttt{Pvalb}
\end{itemize}

These genetically defined lines allow for cell-type-specific recordings, enabling mechanistic comparisons across neuronal classes. For instance, \texttt{Vip} interneurons are known for disinhibitory control and strong directional tuning, while \texttt{Sst} interneurons exhibit broader inhibitory tuning \cite{MillmanEtAl2020}. In this study, we focused on the inhibitory lines (\texttt{Vip}, \texttt{Sst}) due to their high decodability in POYO+ and distinct biological roles in cortical circuits. Recordings were obtained across six visual cortical areas—VISp, VISpm, VISam, VISrl, VISal, and VISl—at depths ranging from 150 to 600 μm, corresponding to cortical layers L2/3, L4, L5, and L6. Each experiment consisted of three hour-long imaging sessions per mouse, with varying combinations of visual area, depth, and cell type, resulting in a rich, multimodal dataset.

\subsubsection{Stimulus Paradigms and Task Selection}

The POYO+ model we interpret in this work was originally trained on neural responses from all five visual stimulus paradigms in the Allen dataset: drifting gratings, static gratings, natural scenes, natural movies, and sparse noise \cite{AzabouEtAl2025}. The SAEs we trained were likewise fit on the full dataset of POYO+ latent outputs. For all mechanistic analyses and interpretability experiments, we restricted our focus to the \textbf{drifting gratings} stimulus task. This canonical paradigm is designed to probe core aspects of visual tuning—including orientation selectivity, direction preference, and temporal frequency encoding—and is widely used in systems neuroscience to characterize receptive fields. Each drifting grating stimulus consists of a full-field sinusoidal pattern drifting perpendicular to its orientation. Stimuli are presented in one of eight directions spaced at $45^\circ$ intervals (0°, 45°, 90°, ..., 315°) and at one of five temporal frequencies (1, 2, 4, 8, and 15 Hz), yielding 40 unique stimulus conditions. Each condition is shown for 2 seconds, followed by a 1-second gray screen (mean luminance), with randomized ordering across trials. Conditions are typically repeated 15 times per session to enable robust trial averaging. In addition, blank sweeps—periods of mean luminance with no stimulus—are intermittently presented to capture baseline and spontaneous activity.

\subsubsection{Filtering for Interpretability}

To reduce anatomical variability and simplify interpretability, we further filtered the data to include only recordings from the \textbf{primary visual cortex (VISp)}. VISp is well-studied and known to contain neurons with sharply tuned responses to drifting gratings, making it a suitable region for evaluating how model representations align with interpretable neurobiological properties such as orientation and temporal frequency tuning. This restriction preserved substantial diversity in Cre-line expression while enabling clearer mechanistic insight into feature-specific model behavior.

\subsection{SAE Hyperparameter Sweep Details}

To optimize the sparse autoencoder (SAE) used in our transformer interpretability experiments, we conducted a hyperparameter sweep using Bayesian optimization, targeting minimization of the average training loss. 
The search configuration is summarized in Table~\ref{tab:sweep}.

\begin{table}[H]
\centering
\caption{Hyperparameter sweep configuration for SAE architecture optimization.}
\label{tab:sweep}
\begin{tabular}{ll}
\toprule
\textbf{Parameter} & \textbf{Search Space} \\
\midrule
Learning rate (lr) & Uniform distribution between 0.0015 and 0.0038 \\
Batch size & Fixed at 16384 \\
Gradient accumulation steps & \{4, 6\} \\
Expansion factor & Fixed at 3 \\
Epochs & \{10, 15, 20\} \\
Use Top-$k$ & True \\
Top-$k$ value & \{15, 20\} \\
Top-$k$ per sample & True \\
\midrule
Optimization method & Bayesian optimization \\
Objective metric & Average training loss (minimized) \\
Run cap & 100 runs \\
Early termination & Hyperband, minimum 3 epochs \\
\bottomrule
\end{tabular}
\end{table}

The sweep was run with early termination via Hyperband to efficiently prune underperforming runs, allowing rapid exploration of the Top-$k$ sparse regime relevant for achieving interpretable latents.

\subsection{SAE Training Configuration}

The sparse autoencoder (SAE) was trained using a custom PyTorch Lightning module. The SAE architecture applies a single linear encoder and decoder with bias terms, ReLU nonlinearity, and optional neuron ablation masking. Decoder weights are L2-normalized to improve stability during reconstruction. Training minimized a composite loss comprising reconstruction error (mean squared error) and a sparsity penalty (L1 norm on latent activations). Training was performed across 4 NVIDIA RTX 3080 GPUs using distributed data parallelism (DDP). 
Key training details are summarized below:

\begin{itemize}
    \item \textbf{Optimizer:} Adam with learning rate selected via Bayesian hyperparameter sweep.
    \item \textbf{Learning Rate Scheduler:} Reduce on plateou on validation loss with a patience of 3 epochs and a minimum learning rate of $10^{-5}$.
    \item \textbf{Loss Terms:} 
    \begin{itemize}
        \item Reconstruction Loss ($\mathcal{L}_{\text{reconstruction}}$): MSE between input and reconstructed output.
    \end{itemize}
    \item \textbf{Model Features:} 
    \begin{itemize}
        \item Optional latent ablation masking during forward passes.
        \item R\textsuperscript{2} score tracking for reconstruction fidelity.
        \item Sparsity Loss ($\mathcal{L}_{\text{sparsity}}$): Sum of absolute activations.
    \end{itemize}
\end{itemize}

The SAE input dimensionality was expanded by a factor of 3, and Top-$k$ sparsity selection was applied on a per-sample basis.

\subsection{Decoding Cre‐Line and Brain‐Region Labels}
\label{appendix:cre_decoding}

\subsubsection{Data Preparation and Latent Extraction}

For each neural time window from the drifting-gratings stimulus task, we extracted latent vectors from two representations:
\begin{itemize}
  \item \textbf{Baseline transformer latents} ($d = 64$): the final-layer latents from POYO+.
  \item \textbf{SAE latents} ($d = 192$): activations of the TopK sparse autoencoder applied to the baseline latents.
\end{itemize}
This yielded two matrices:
\[
X_{\mathrm{dec}} \in \mathbb{R}^{N \times 64}, \quad X_{\mathrm{SAE}} \in \mathbb{R}^{N \times 192},
\]
where $N$ is the number of neural samples, each corresponding to a specific time point in the experiment. Associated labels $y_i$ denote either the genetic Cre-line or brain region for sample $i$. To focus on interpretable comparisons, all analyses were restricted to samples from the primary visual cortex (VISp).

\subsubsection{Class Balancing and Preprocessing}

Cre-line and brain region labels were typically imbalanced, with certain classes dominating. To address this, we downsampled all classes to the size of the smallest class. This ensured a balanced classification problem and prevented accuracy inflation from dominant labels. All features in $X$ were standardized to have zero mean and unit variance across samples before model training.

\subsubsection{Logistic Regression and Cross-Validation}

We used a multinomial logistic regression model with an $\ell_2$ penalty (regularization strength $C=1$), trained using the \texttt{lbfgs} solver and a maximum of 2000 iterations. We evaluated model generalization with 3-fold cross-validation using a GroupKFold scheme, ensuring that samples from the same recording session were not split across folds. This avoids data leakage across animals and accounts for session-specific variability.

\subsubsection{Statistical Assessment of Latent-Class Associations via Bootstrapping}
\label{appendix:decoding_stats}

To determine whether a given latent unit significantly encodes a specific label, we performed a coefficient stability analysis using non-parametric bootstrapping. This procedure tests whether the association between a latent feature and a cell type is consistently preserved across resampled training sets, and whether the sign of the learned weight is statistically stable.

Let $\mathcal{D} = \{(\mathbf{x}_i, y_i)\}_{i=1}^{N}$ denote the dataset, where $\mathbf{x}_i \in \mathbb{R}^d$ is the vector of latent activations for sample $i$, and $y_i$ is the corresponding cell type label. We fit a one-vs-rest logistic regression classifier $f_\theta$ on $\mathcal{D}$, yielding class-specific weight vectors $\theta_c \in \mathbb{R}^d$ for each cell type $c$.

To estimate the stability of each coefficient $\theta_{cj}$ (i.e., the weight assigned to latent unit $j$ for class $c$), we apply the following procedure:

\begin{enumerate}
    \item Generate $B$ bootstrap samples $\mathcal{D}^{(1)}, \dots, \mathcal{D}^{(B)}$ by sampling $N$ datapoints with replacement from $\mathcal{D}$.
    \item For each bootstrap sample $\mathcal{D}^{(b)}$, train a logistic regression classifier to obtain class-specific coefficients $\theta_c^{(b)}$.
    \item For each latent dimension $j$ and class $c$, collect the bootstrapped coefficient values:
    \[
        \mathcal{C}_{cj} = \left\{ \theta_{cj}^{(b)} \right\}_{b=1}^B.
    \]
    \item Compute the empirical mean:
    \[
        \bar{\theta}_{cj} = \frac{1}{B} \sum_{b=1}^{B} \theta_{cj}^{(b)}.
    \]
    \item Compute a $95\%$ confidence interval:
    \[
        \text{CI}_{cj}^{95\%} = \left[ \text{Percentile}_{2.5}(\mathcal{C}_{cj}), \text{Percentile}_{97.5}(\mathcal{C}_{cj}) \right].
    \]
    \item Estimate a two-sided $p$-value by calculating the proportion of bootstrap samples whose sign disagrees with the mean:
    \[
        p_{cj} = \frac{1}{B} \sum_{b=1}^{B} \mathbb{I}\left[ \theta_{cj}^{(b)} \cdot \bar{\theta}_{cj} < 0 \right].
    \]
\end{enumerate}

\subsubsection{Latent Significance Criterion}

We define a latent unit $j$ as significantly associated with class $c$ if:
\[
    p_{cj} < \alpha,
\]
where $\alpha = 0.05$ is a user-specified significance threshold. This criterion implies that the coefficient's sign is consistent across resampled datasets with at least $95\%$ confidence. These significant latents are further categorized by sign (positive or negative) and latent type (e.g., orientation or temporal frequency).

\subsubsection{Evaluation and Visualization}

We report the mean cross-validation accuracy, per-class accuracy, and normalized confusion matrices for each latent type (baseline vs.\ SAE). We also visualized latent structure using UMAP, projecting the high-dimensional latent activations into 2D to examine whether SAE latents exhibit greater class separability. In some cases, specific Cre-line labels (e.g., VIP, SST) were highlighted to assess disentanglement of biologically meaningful structure. All analyses were implemented in Python using \texttt{scikit-learn} for logistic regression, \texttt{GroupKFold} for validation, and \texttt{umap-learn} for dimensionality reduction. 

\subsection{Unsupervised Clustering of Latent Receptive Fields}
\label{sec:unsupervised_clustering}

To investigate whether SAE latent dimensions encode distinct or overlapping functional properties, we performed an unsupervised clustering analysis based on their activation patterns across drifting gratings stimuli. Each SAE latent was first mapped to an activation matrix $A_{i,j} \in \mathbb{R}^{5\times8}$ (temporal frequency $\times$ orientation) as described in the methods. 
These matrices can be interpreted as "receptive fields" summarizing each latent's selectivity to combinations of temporal frequency and grating orientation. We then applied the following unsupervised clustering pipeline:

\begin{enumerate}
    \item \textbf{Preprocessing:} 
    Latents with negligible overall activation (sum of absolute values $<0.1$) were excluded to focus analysis on meaningful units.
    Each $A_{i,j}$ matrix was flattened into a 40-dimensional vector and standardized via z-scoring.
    
    \item \textbf{Dimensionality Reduction:} 
    Uniform Manifold Approximation and Projection (UMAP) was applied to reduce the activation vectors to two dimensions. 
    UMAP parameters were set to $n_{\text{neighbors}}=15$, $\text{min\_dist}=0.1$, using correlation distance as the metric.
    
    \item \textbf{Clustering:} 
    HDBSCAN was then run on the 2D UMAP embeddings to discover clusters of latents with similar receptive fields.
    HDBSCAN was configured with a minimum cluster size of 5 and minimum samples of 1, using Euclidean distance.
    
    \item \textbf{Evaluation and Visualization:} 
    Cluster quality was assessed using silhouette scores when multiple clusters were found.
    We visualized both the UMAP embeddings colored by cluster identity and summarized the average activation patterns within each cluster.
\end{enumerate}

This analysis provided an unsupervised grouping of latent dimensions into distinct functional classes based purely on their response profiles, without using behavioural or genetic labels. The clustering revealed that many SAE latents spontaneously grouped into interpretable categories with distinct orientation and temporal frequency selectivity, highlighting the emergence of specialized representations even without explicit supervision.

\end{document}